\documentstyle[aps]{revtex}
\begin{document}
\title{Generic criticality in a model of evolution\\}
\author{Adam  Lipowski \cite{byline}\\}
\address{Department of Physics, A. Mickiewicz University, Ul. Umultowska 85, 
61-614 Pozna\'{n}, Poland}
\date{\today}
\maketitle
\begin{abstract}
Using Monte Carlo simulations, we show that for a certain model of biological 
evolution, which is driven by non-extremal dynamics, active and absorbing 
phases are separated by a critical phase.
In this phase both the density of active sites $\rho(t)$ and the survival 
probability of spreading $P(t)$ decay as $t^{-\delta}$, where 
$\delta \sim 0.5$.
At the critical point, which separates the active and critical phases, 
$\delta\sim 0.29$, which suggests that this point belongs to the so-called 
parity-conserving universality class.
The model has infinitely many absorbing states and, except for a single point, 
has no conservation law.
\end{abstract}
Recently statistical mechanics of complex systems attracts a lot of attention.
The main motivation of studying these systems is the belief 
that their qualitative understanding can be obtained by studying relatively 
simple mathematical models.
Indeed, there is a number of examples where (spin-)glasses, proteins, 
biological evolution, societies or economies were described in terms of very 
simple models~\cite{MEZARD,BAK}.
In quantitative terms, complexity is very often related with the absence of
a characteristic length- or time scale (i.e., scale invariance).
For example, paleontological data suggest 
that outbreaks of evolutionary activity substantially varied in sizes and in 
addition were correlated over large periods of time~\cite{SOLE}.
Similar features seem to characterize other, at first sight unrelated, 
processes like earthquakes, stock-exchange fluctuations or flow of sand.

The absence of characteristic scales is a well-known property of critical 
systems in the field of equilibrium statistical mechanics.
However, in the equilibrium statistical mechanics, criticality is an 
exception rather than a rule and it requires fine-tunning of control 
parameter(s).
On the other hand, the apparent abundance of scale-invariance among complex 
systems suggests that this property should be in some sense generic and 
should not require such a fine tunning.

An interesting idea, which tries to explain the scale invariance 
in various systems, was proposed by Bak, Tang and Wiesenfeld under the name of 
self-organized criticality (SOC)~\cite{BTW}.
They have shown that dynamics of some simple systems might naturally lead 
these systems to the critical state.
Such a behaviour was subsequently observed in a number of other models.

However, models exhibiting SOC are usually driven by very special dynamics.
This is either the so-called extremal dynamics~\cite{MAYA} (which drives some
evolutionary models) or conservative 
dynamics (which drives sandpile-like models).
When these dynamical rules are even slightly violated, the 
criticality is usually destroyed.
For example, in the Bak-Sneppen model ~\cite{BAKSNEPPEN} describing the 
evolution of an ecosystem, the criticality is lost when we modify the rule 
that only the least-fitted species dies out and is replaced by a new one.
However, on biological grounds, one expects that extinction might happen 
to a better-fitted species as well.
It would be desirable to construct a model which would not be driven by 
such a special dynamics but whose criticality would be in some sense generic.

In search of the generic criticality, we might recall that such models exist in
equilibrium statistical mechanics, and the prime example is the XY model.
In this model the low-temperature phase is critical and correlation
functions decay in a power-law way.
Above certain temperature the criticality of the model is destroyed and the 
system is in a disordered phase, where correlation functions decay 
exponentially~\cite{KT}.

Despite the wealth of models with critical behaviour, the existence of
generic criticality in the nonequilibrium statistical mechanics 
is still an open problem~\cite{GRINSTEIN}.
In some cases, certain symmetries, conservation laws or separation of 
time scales are responsible for the criticality of the system.
These factors play also an important role in SOC models.
Recently, some SOC models were related with more general 
models~\cite{SORNETTE,VESP98,VESP00}.
It turns out that in some cases the SOC corresponds to the critical points of 
these more general models.
However, the criticality of the latter models most likely resembles an 
ordinary criticality (i.e., the criticality exists only at some isolated 
points).

In the present paper we study a model of biological evolution.
The model describes an ecosystem at the coarse-grained level similarly to the 
Bak-Sneppen model, but it is driven by non-extremal dynamics.
We show that in a certain range of a control parameter certain quantities 
exhibit power-law behaviour and thus the model might be said to be generically 
critical.
The behaviour of our model is a consequence of a certain symmetry, which 
places the model in the so-called parity-conserving (PC) universality 
class.

Our model is a variant of other recently introduced 
models~\cite{LIPLOP,LIP99,LIP00}.
It is defined on a one-dimensional lattice, where for each 
bond between the nearest-neighbouring sites $i$ and $i+1$, we introduce bond 
variables $w_{i,i+1}\in (-0.5,0.5)$.
Introducing the parameter $r$, we call the site $i$ active when 
\begin{equation}
w_{i,i+1}w_{i-1,i}<r.
\label{e1}
\end{equation}
Otherwise, the site is called nonactive.
The model is driven by random sequential dynamics and when the active site 
$i$ is selected, we assign anew, with uniform probability,  two bond 
variables $w_{i,i+1}$ and $w_{i-1,i}$.
Nonactive sites are not updated, but updating a certain (active) site might
change the status of its neighbours.
The above rules immediately imply the existence of an absorbing state, i.e., 
the one without active sites.

One can interpret the sites as species with the fitness being a product of 
the attached bond variables.
When the fitness of a certain species is lower than a threshold value 
$r$, the species becomes extinct and is replaced by another species.

To examine the properties of our model, we used Monte Carlo simulations.
Since the implementation of the above dynamical rules on a computer is 
straightforward, below we present only the results of these simulations.
An important quantity characterizing this model is the steady-state density 
of active sites $\rho$.
Fig.~\ref{f1} shows the density $\rho$ as a function of $r$.
The simulations were performed for the linear system size $L=10^5$ and we
checked that the presented results are, within a small statistical error, 
size-independent.
For each $r$, after relaxing the random initial configuration for 
$t_{{\rm rel}}=10^4$, we made measurements during runs of $t=10^5$ 
(the unit of time is defined as a single on average update/lattice site).
This figure suggests that the model undergoes 
a continuous transition at $r=r_{{\rm c}} \sim 0.027$ and for $r<r_{{\rm c}}$ 
the model should be in the absorbing phase with $\rho=0$.
The simulations close to the critical point ($r<0.03$) were more extensive 
and we used $L=5\cdot 10^5$, $t_{{\rm rel.}}=5\cdot 10^4$ and $t=10^6$.
Let us also notice that since bond variables are continuous, there is 
a continuous degeneracy of the absorbing state.

We also observed that for $r<r_{{\rm c}}$ the system is rather reluctant
in reaching an absorbing state.
Only for $r<0$, the system reaches such a state quite quickly.
Results of our simulations presented below show that for 
$0<r<r_{{\rm c}}$ the model remains in the critical phase.

First, we examined the time evolution of the density of active sites $\rho(t)$.
After the system starts from an arbitrary initial configuration, containing a 
finite fraction of active site, $\rho(t)$ should decay to zero in the 
absorbing phase and this decay should be faster than the power law.
At the critical point one expects a power-law decay $\rho(t)\sim t^{-\delta}$.
In the active phase $\rho(t)$ asymptotically remains positive.

In Fig.~\ref{f2} we show $\rho(t)$ as a function of $t$ plotted in the 
logarithmic scale.
For $r=0.03$ the density $\rho(t)$ clearly approaches a positive value, which
confirms that in this case the model is in the active phase.
Moreover, a faster than power-law decay is observed for $r=-10^{-4}$ and 
$-10^{-6}$.
However, for $r=0$ and 0.02 our simulations show a power-law 
decay with the exponent $\delta=0.50(1)$.
For $r=0.027$, which is very close to the critical point (see
Fig.~\ref{f1}), the exponent $\delta=0.29(1)$.

Critical properties of models with absorbing states can be also studied using 
the so-called dynamic (or epidemic) method~\cite{TORRE}.
In this method, we prepare the system in one of the absorbing states except 
a localized (usually at a single site) activity.
Subsequently, the system evolves according to its dynamical rules and and we 
monitor statistical properties of such runs.
One of the important quantities in this technique is a probability $P(t)$ 
that a given activity survives until time $t$.
On general grounds, one expects that $P(t)$ behaves similarly to $\rho(t)$.
Namely, in the active phase $P(t)$ tends to a finite value, in the absorbing 
phase it rapidly (faster than a power law) decreases to zero and in the 
critical phase it decays as $t^{-\delta'}$, where $\delta'$ is an exponent, 
which in general might be different than $\delta$~\cite{MENDES,MARQUES}.

Fig.~\ref{f3} shows the results of our dynamical simulations.
Simulations were performed for sizes which ensured that the spreading 
activity did not reach the boundaries of the lattice (typically 
$L=2\cdot 10^4$ is sufficient).
The number of runs varied from $10^6$ for $r=-10^{-4},-10^{-6}$ and 0 
to $2\cdot 10^4$ for $r=0.03$
The results shown in Fig.~\ref{f3} lead to the same conclusions as those in 
Fig.~\ref{f2}: for $0\leq r<r_{{\rm c}}$ the model remains in the critical 
phase
with $\delta'=0.50(1)$.
For $r<0$ the survival probability decays most likely faster than
an inverse power of $t$.

The fact that for $r<0$ the model is in the absorbing phase is to some 
extent understood.
Namely, for $r<0$ there exists a finite probability that after updating a 
pair of sites will become nonactive forever.
Indeed, when one of the anew selected bonds (say, $w_{i,i+1}$) satisfies the 
condition
\begin{equation}
|w_{i,i+1}|<-r/(0.5),
\label{e2}
\end{equation}
then the sites $i$ and $i+1$ become permanently nonactive.
Namely, no matter what are
the other bonds (i.e., $w_{i-1,i}$ and $w_{i+1,i+2}$) attached to these 
sites, they will always remain nonactive.
For $r<0$ there is a finite probability of satisfying Eq.~\ref{e2} and the 
above  mechanism leads to the rapid decrease of active sites and hence 
the system quickly reaches an absorbing state.
The above mechanism is not effective for $r\geq 0$ since there is no value 
which would ensure permanent nonactivity of a certain site.
Basically the same mechanism is at work in another model with 
absorbing states~\cite{LIP00}.

More detailed analysis of $\rho$ in the vicinity of 
the critical point suggests that the exponent $\beta$ is  
slightly less then unity.
Together with the value of $\delta\sim 0.29$  at the critical point 
$r=r_{{\rm c}}\sim 0.027$ (as estimated from the results in Fig.~\ref{f2}), 
this strongly suggests that this model behaves 
similarly to some other models, which are commonly termed as the 
parity-conserving universality class~\cite{GRASS84,HAYE97}.
In this class of models, it is already known that in the critical phase
$\delta=0.5$.
An interesting feature of our model is the fact that the critical 
phase terminates at a certain point ($r=0$) and (most likely) an exponential 
decay sets in.
Let us also notice that for $r=0.027$ the slope in Fig.~\ref{f3} is also 
close to $0.29$.
However, this might be a coincidence, since some variability of this exponent
in dynamic Monte Carlo method is an anticipated feature~\cite{MENDES,MARQUES}.

Initially, the behaviour of models belonging to the PC universality class was 
thought to be determined by the local conservation laws~\cite{GRASS84}.
Later, however, some models were found which do not possess this property but 
which exhibit the same critical behaviour\cite{HAYE97}.
Since these models have a symmetric and double-degenerate absorbing state, it 
seemed that this is another property which leads to the PC criticality.
Recently, a model with infinitely many absorbing states was found, which also 
belongs to the PC universality class but this might be again attributed to 
some conservation law in its dynamics~\cite{MARQUES}.
The present model has an infinitely-degenerate absorbing state and no 
conservation law.
We attribute the PC criticality of this model to a certain global symmetry of 
this model.
Namely, one can easily see that inverting all bond variables 
($w_{i,i+1}\rightarrow -w_{i,i+1}$) one does not change (non-)activity of 
any site.
This Ising-like symmetry determines the structure of any absorbing state 
for $r\geq0$: all bond variables must be either positive or negative.

Let us note that although there is no conservation law in this model in 
general, there is such a law for $r=0$.
Indeed, for $r=0$ the dynamics of the model is special: it is only the sign of 
$w_{i,i+1}$ which matters and only those sites are active where negative and 
positive $w_{i,i+1}$ meet.
As a result, the dynamics of the model is equivalent to a certain branching 
annihilating random walk (BARW) with an even number of offsprings~\cite{COMM1}.
The power-law characteristics $t^{-0.5}$ are already known for related BARW 
models~\cite{CARDY}.

In summary, we have shown that a certain 
evolutionary model with non-extremal dynamics exhibits generic criticality.
Such a behaviour is most likely related with a special symmetry of this model.
One might hope that this criticality is to some extent robust with respect to 
structural perturbations of this model (other lattices, definitions
of fitness function, etc.,).
Since for some of these variants the analogy with random walk models might 
not hold, it is possible that the criticality of such models will exhibit 
some sort of non-universality.
However, analysis of such extensions is left as a future problem.
\acknowledgements
I thank Dr.~H.~Hinrichsen for interesting discussion and the Department of 
Mathematics of the Heriot-Watt University (Edinburgh, Scotland) for 
allocation of the computer time.

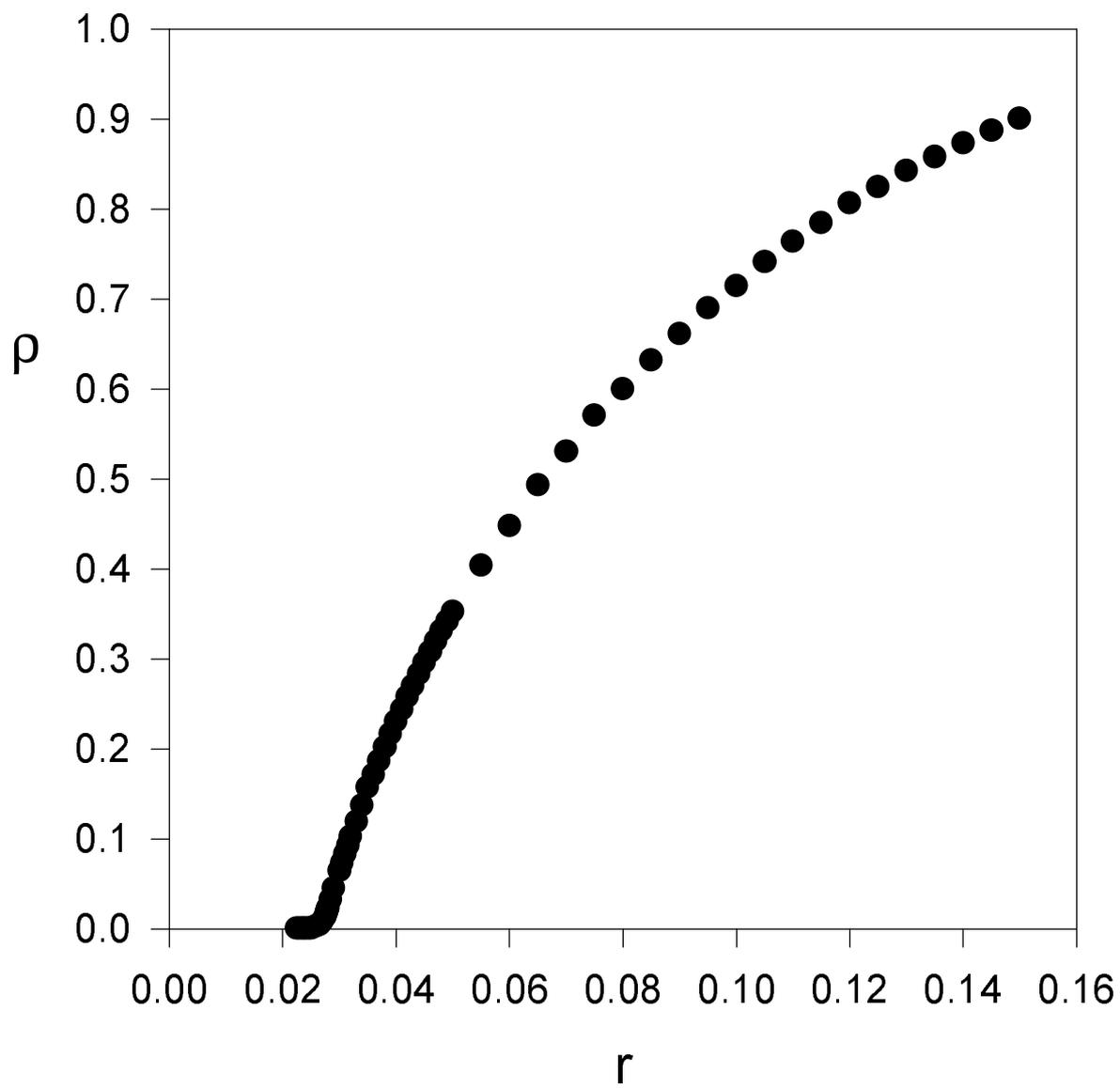
\begin{figure}
\caption{The steady-state density of active sites $\rho$ as a function of $r$.}
\label{f1}
\end{figure}
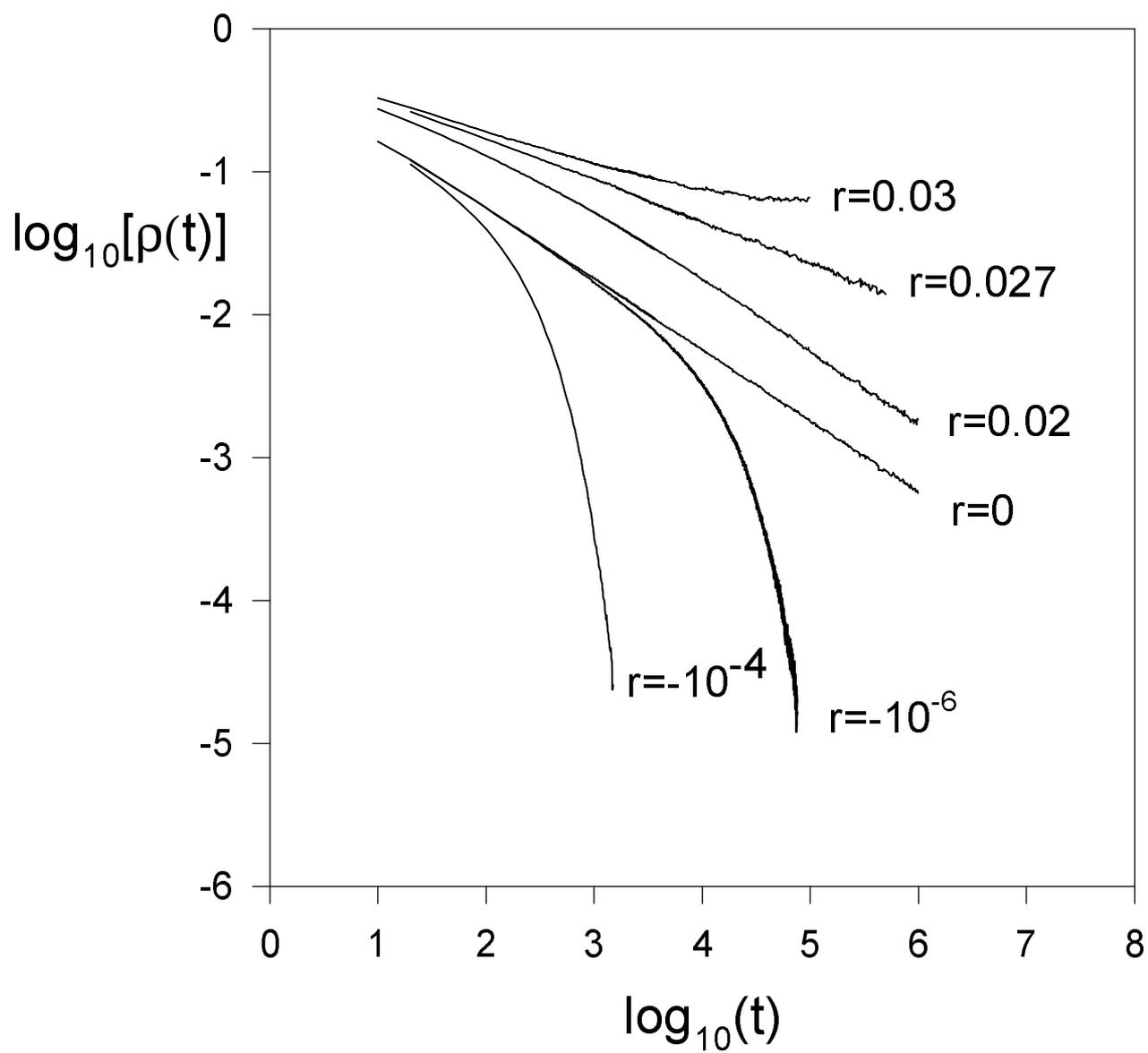
\begin{figure}
\caption{The density of active sites $\rho(t)$ as a function of $t$ plotted 
in the log-log scale.
Simulations were made for $L=10^5$ and for each $r$ we averaged over about 50
independent runs.}
\label{f2}
\end{figure}
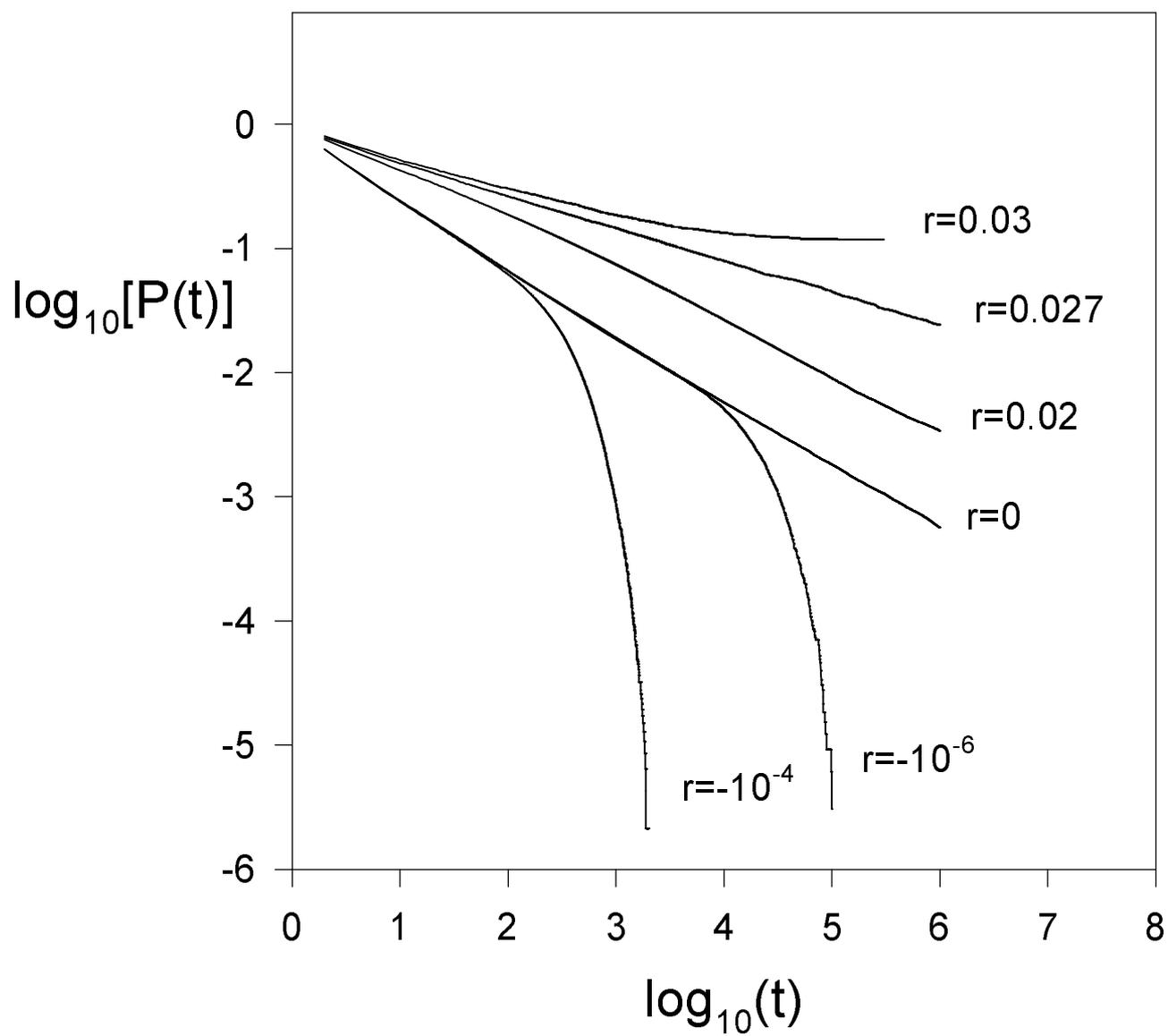
\begin{figure}
\caption{The survival probability $P(t)$ as a function of $t$ plotted 
in the log-log scale.
Initially we set $w_{i,i+1}=w_0=0.2$ for all bonds except for those 
surrounding a certain site which was set as active.
Different choice of $w_0$ might change the asymptotic slope at 
$r=r_{{\rm c}}\sim 0.027$, but it should not affect the asymptotic $t^{-0.5}$
decay for $0\leq r<r_{{\rm c}}$.
}
\label{f3}
\end{figure}

\begin{references}
\bibitem{COMM1} One can easily see, that for $r=0$ the number of active sites 
in the system after updating an active site changes by 0 or 2.
\end{references}
\end{document}